\documentclass{acm_proc_article-sp}
\pdfoutput=1

\begin{document}

\title{Let's Annotate to Let Our Code Run in Parallel}
%
%
%
%
%

\numberofauthors{1} 
%
\author{
%
%
\alignauthor
Patrizio Dazzi\\
       \affaddr{ISTI - CNR}\\
       \email{patrizio.dazzi@isti.cnr.it}
}

\date{5 June 2013}

\maketitle
\begin{abstract}

This paper presents an approach that exploits Java annotations to provide meta information needed to automatically transform plain Java programs into parallel code that can be run on multicore workstation. Programmers just need to decorate the methods that will eventually be executed in parallel with standard Java annotations. Annotations are automatically processed at launch-time and parallel byte code is derived. Once in execution the program automatically retrieves the information about the executing platform and evaluates the information specified inside the annotations to transform the byte-code into a semantically equivalent multithreaded version, depending on the target architecture features. The results returned by the annotated methods, when invoked, are futures with a wait-by-necessity semantics.

\end{abstract}


\section{Introduction}

Developing parallel applications is, in general, much more complex than developing sequential applications. Besides being in charge of the whole parallel application structure, programmers have to deal with communications, synchronization, mapping and scheduling structure. As the programmers usually write applications directly interacting with the middleware, 
the whole process is cumbersome and error prone.
So far, several efforts have been spent to face this problem, and several approaches have been conceived to design high-level programming languages/environments that can automate most of the tasks required to implement working and efficient parallel applications.

Other approaches offer a lower abstraction level but allow more programming freedom and guarantee a higher level of personalization.
In other words, programmers can customize their applications and deal with some aspects related to the parallelism as, for example, parallelism degree and the parallel program structure. 

The approaches belonging to this category force the programmer to structure the parallel application he wants to implement adequately. Typically, such approaches allow the application ``business logic'' to be separated from the activities required to coordinate and to synchronize parallel processes \cite{qosassist}.
On the other side, several environments have been proposed to use more classical, low level programming paradigms. However, all these approaches, while leaving the programmer a higher freedom of structuring the parallel applications in an arbitrary way, require the programmers explicitly deal with all the awkward details mentioned above.

In this work, we describe Parallel Abstraction Layer (PAL), originally presented in~\cite{danelutto2008pal}. It aims at avoiding the problems typically present in a fully automated parallel approach \cite{160438}, PAL leaves to programmer the responsibility to choose which parts of code have to be computed in parallel through the insertion of non-functional requirements in the source program code.
Using the information provided by programmers PAL transforms the program code into a parallel one.

\section{Parallel Abstraction Layer}\label{sec:idea}
PAL is an approach conceived around a quite simple but very embraceable, well-known, opinion ``\emph{...people know the application domain and can better decompose the problem, compilers can better manage data dependence and synchronization}'' \cite{grimshaw93mentat}. the PAL approach to parallel programming fundamentally relies on programmer knowledge to properly ``structure'' the parallel schema of an application and then let to the compiler/run time tool ability to efficiently implement such schema. 

Basically, this almost matches the algorithmic skeletons approach \cite{ColeCloset}. 
PAL represents a general-purpose mechanism based on very simple applications structuring. In fact the programmer is only required to specify some hints that are exploited by the runtime support to implement a parallel version of the application code. 
These hints are specified through the annotation mechanisms provided by Java~\cite{javaAnnotation}.

The programmers are required to give some kind of ``parallel structure'' to the code directly at the source code level, as it happens in the algorithmic skeleton case. However, the approach discussed in this work presents at least two additional advantages.
\begin{itemize}
\item
First, annotations can be ignored and the semantics of the original sequential code is preserved. This means that the programmer application code can be run through a classical compiler/interpreter suite and debugged using normal debugging tools.
\item
Second, annotations are processed at load time, typically exploiting reflection properties of the hosting language. As a consequence, while handling annotations, a bunch of knowledge can be exploited which is not available at compile time (e.g. running machines) and this can lead to more efficient parallel implementations of the user application.
\end{itemize}

In order to experiment the feasibility of the proposed approach, we considered the languages that natively support code annotations for developing a validation prototype.
Both Java and .NET frameworks provide an annotation mechanism. They also provide an intermediate language (IL), portable among different computer architecture (compile once -- run everywhere), and holding some information typically only available at source code level (e.g. code annotations) that can be used in the runtime for optimization purposes.

The transformation process is done at load time, namely the time when we have all the information needed to optimize the restructuring process with respect to the available underlying resources. The code transformation works at IL level thus it does not need that the application source code is sent on target architecture. Furthermore, IL transformation introduces in general fewer overheads than the source code transformations followed by re-compilation.

PAL transforms the annotated code in a parallel fashion by asynchronously executing parts of the original code. The parts to be executed asynchronously are individuated by the user annotations.  In particular, we used Java and therefore the more natural choice was to individuate method calls as the parts to be asynchronously executed. PAL translates the IL codes of the ``parallel'' part by structuring them according with the features of the target architecture.
Asynchronous execution of method code is achieved by exploiting the concept of \emph{future} \cite{caromel05theory, caromel04asynchronous}. When a method is called asynchronously it immediately returns a future, that is a stub ``empty'' object. The caller can then go on with its own computations and use the future object just when the method call return value is actually needed. If in the meanwhile the return value has already been computed, the call to reify the future succeeds immediately, otherwise it blocks until the actual return value is computed and then returns it.

PAL programmers have just to put a \verb1@Parallel1 annotation on the line right before method declaration to mark that method as a candidate for asynchronous execution. This allows keeping applications similar to normal sequential applications, actually. Programmers may simply run the application through standard Java tools to verify it is functionally correct. The PAL approach also avoids the proliferation of source files and classes, as it works transforming IL code, but raises several problems related to data sharing management. As an example, methods annotated with a  \verb1@Parallel1 cannot access class fields: they can only access their own parameters and the local method variables. This is due to the impossibility to intercept all the accesses to the class fields, actually.
Then, PAL automatically performs at load time the activities aimed at achieving the asynchronous and parallel execution of the PAL-annotated methods and at managing any consistency related problems, without any further programmer intervention.

\section{The PAL prototype}\label{sec:pro}
To validate our approach, we implemented a PAL prototype in Java, as it provides a manageable intermediate language (Java byte-code \cite{javaVMSpec}) and natively supports code annotations. s 
The prototype works taking the program byte-code as input and transforming it in a parallel byte-code. In order to do this it uses ASM \cite{bruneton02asm}: a Java byte-code manipulation framework.

The current prototype accepts only one kind of attribute to the \verb1@Parallel1 annotation: a \texttt{parDegree} denoting the maximum number of processing elements to be used for the method execution. PAL uses such information to make a choice between the multithreaded and 
distributed version. This choice is driven by the number of processors/cores available on the host machine: if the machine owns a sufficient number of processors the annotated byte-code directly compiled from user code is transformed in a semantically equivalent multithreaded version. 

In order to enable the PAL features, the programmer has only to add a few lines of code. As an example consider a program computing the Mandelbrot set. The \texttt{Mandelbrot} class uses a \texttt{@Parallel} annotation to state that all the input data (e.g. \texttt{createLines} calls) should be computed in parallel, with a specified parallelism degree. Unfortunately, due to some Java limitations, the programmer must specify an ad-hoc return type (\texttt{PFFuture}), and consequently return an object of this type. \texttt{PFFuture} is a template defined by the PAL framework. It represents a container needed to enable the future mechanism. The type specified as argument is the original method return type. Initially, we tried to have to a more transparent mechanism for the future implementation, without any explicit Future declaration. It consisted in the load-time substitution of the return type with a PAL-type inheriting from the original one. In our idea, the PAL-type would have filtered any original type dereferentiation following the \emph{wait-by-necessity} \cite{caromel89wait} semantics. Unfortunately, we had to face two Java limitations that limit the current prototype to the current solution. 

These limitations regard the impossibility to extend some widely used Java BCL classes (String, Integer,...) because they are declared \texttt{final}, and the impossibility to intercept all class field accesses.

In the \texttt{Main} \ class, the user just asks to transform the \texttt{Main}\ and the \texttt{Mandelbrot} \ classes with PAL, that is, to process the relevant PAL annotations and to produce an executable IL which exploits parallelism according to the features (hw and sw) of the target architecture where the \texttt{Main}\ itself is being run.

\section{Related work}\label{sec:rwork}
PAL offers a simple yet expressive technique for parallel programming. By exploiting ``runtime compilation'' it adapts the executable code to different architectures. It does not introduce a new or different paradigm, while exploiting parallelism at the method call level. So far have been proposed a certain number of systems based on similar ideas.
However, although different experiments exist in the so-called concurrent object-oriented languages scenario (COOLs) \cite{philippsen00survey}, we decided to discuss only those actually very similar to PAL.
%
%
In \cite{JavaOpenMP}\ the authors propose a Java version of OpenMP giving to the programmers the possibility to specify some PRAGMAs inside comments to source code. These pragmas are eventually used by a specific java HPC compiler to transform the original program in a different one exploiting parallelism, for instance through loop-parallelization.
There are three important differences between this approach and the ours one: first of all PAL works at method level making method invocations asynchronous, while the work presented by Klemm et al. mainly works at the loop-parallelization level. Another very important difference is related to the moment in which the transformation is made: this approach works at compile time starting from source-code, while PAL directly transforms the byte-code at load and run time. As a consequence, PAL may optimize its transformation choices exploiting the knowledge available on the features of the computing resources of the target execution platform.
Eventually, PAL uses \textit{java Annotations} to enrich the source code, instead the Java version of OpenMP uses the source code comments. 
The former approach exploits Java basic features, in particular annotations, which type and syntax are checked by compiler, with the limitation that annotations cannot be placed everywhere in the source code. the latter solution instead is more ``artificial'' but it is not limited to classes, methods and class fields (as the \textit{java Annotations}) and it can be also applied to pure Java code blocks.
If we limit the discussion to the approaches that transform a sequential object-oriented program into a concurrent one by replacing method invocations with asynchronous calls, (where parallelism can be easily extracted from sequential code without modification, without changing the sequential semantics and the wait for return values can be postponed to the next usage, eventually using future objects) the number of approaches similar to PAL is small. However, some other approaches share single points/features with our PAL approach. 
Java made popular the remote method invocation (RMI) for interaction between objects in disjoint memories. The same properties that apply for parallelizing sequential local calls apply for remote ones, with the advantage that remote calls do not rely on shared memory. Parallelizing RMIs scales much better than local calls, as the number of local processors does not limit the number of parallel tasks. This led to many implementations of asynchronous RMIs.
ProActive is a popular object oriented distributed programming environment supporting asynchronous RMIs \cite{proactive}. It offers a primitive class that should be extended to create remote callable active objects, as well as a runtime system to remotely instantiate this type of objects. Any call to an active object is done asynchronously, and values are returned using future objects. Compilation is completely standard, but instantiation must be done supplying the new object location. All active objects must descend from the primitive active object class, so existing code must be completely encapsulated to become active, as there is no multiple inheritance in Java. Although concurrency is available through asynchronous calls, scalable parallelism is obtained creating several distributed objects, instead of calling several concurrent methods, which is not always a natural way of structuring the parallelism.
Some other systems, at different levels, offer asynchronous remote method calls, like JavaParty~\cite{philippsen97javaparty}, JJPF~\cite{danelutto2005java}, Muskel~\cite{aldinucci2001muskel, danelutto2006joint, danelutto2006joint2} and Ibis\ \cite{Nieuwpoort2005a}. They provide a lower level of abstraction with respect to PAL,  being more concerned with the performance of RMI and efficient implementation of asynchronous mechanisms. Usually they offer a good replacement for the original RMI system, either simplifying object declaration or speeding up the communication. Both rely on specific compilers to generate code, although Ibis generate standard JVM byte-code that could therefore be executed on any standard JVM.

\section{Experimental results}
\label{sec:test} 

To validate our approach we ran some experiments with the current prototype. We conducted our tests on a hyper-threaded bi-processors workstation (Intel Xeon 2Ghz, Linux kernel 2.6). 

\begin{figure}
    \centering
\includegraphics[width=\columnwidth]{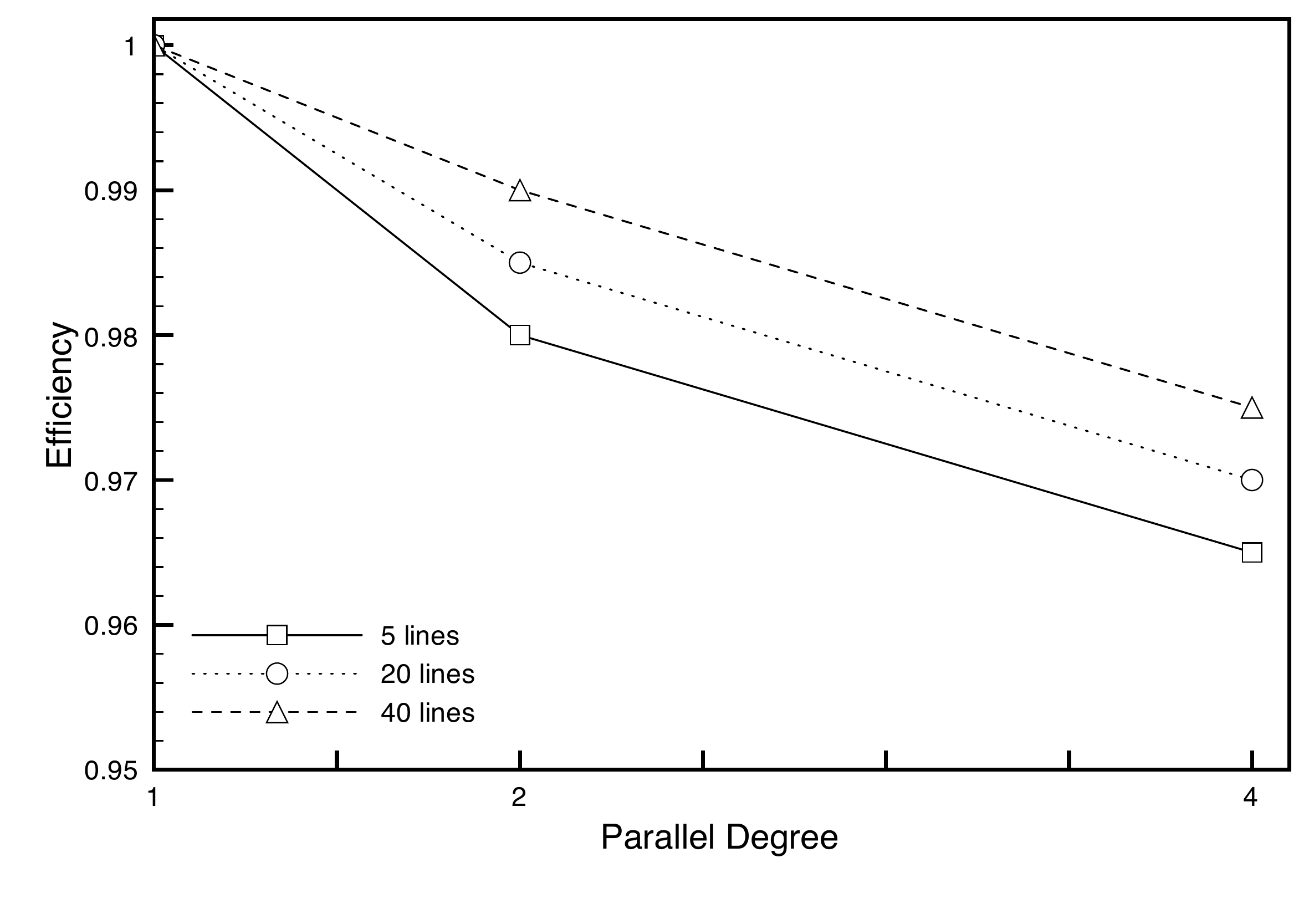}
    \caption{Mandelbrot computation: efficiency comparison with different image resolution, processing element number and task computational weight.}
    \label{fig:efficiency}
\end{figure}

Our test application is a fractal image generator, which computes sections of the Mandelbrot set.

We picked up Mandelbrot set computation as it is a very popular benchmark for embarrassingly parallel computation. PAL addresses exactly these kinds of computations. Most of times, the implementation of these applications requires a significant programming effort, despite being ``easy'' embarrassingly parallel, far more consistent than the effort required to execute the same kind of application exploiting PAL. 
To study in more detail the behavior of the transformed version in several contexts, we ran the fractal generator setting different combinations of resolution (600x400, 1200x800, 2400x1600) and task computational weights, starting from 5 up to 40 lines at time. Clearly when the task size (number of lines to compute) increases, the total number of tasks decreases.

\section{Conclusion and future work} \label{sec:conclusions}
In this paper we present PAL, an approach for easing multicore SMPD programming.  PAL exploits the programmer knowledge provided through annotations to restructure Java programs and make them parallel. The whole process is driven by the analysis of the degree of parallelism specified through annotations. This process is executed at launch time, directly at intermediate language level. This allows obtaining and to exploit at the right time all the information needed to parallelize the applications with respect to the parallel tools available on the target execution environment and to the user supplied non-functional requirements. A load time transformation allows hiding most of parallelization issues.
To validate the approach we developed a PAL prototype that we used it to conduct some preliminary experiments. The results are encouraging and show that the overhead introduced by PAL is negligible, while keeping the programmer effort to parallelize the code low. Anyway, the prototype we developed presents  some limitations. Basically, the class fields are not accessible from PAL-annotated methods, moreover, the programmer has to include an explicit dereferentiation of objects returned by PAL-annotated methods.
In the next future we plan to refine the implementation to address some of these issues as well as extending the approach to be useful also in distributed architectures like Grids or Cloud (and Federation of Clouds too \cite{carlini2012cloud, coppola2012contrail}).
We think this will be interesting and will support cross-fertilization between these concepts as happened in~\cite{aldinucci2008behavioural}.

\bibliographystyle{abbrv}
\bibliography{sigproc}  

\balancecolumns
\end{document}